# An Italian Gender Equality Index

**Lorenzo Panebianco**[1]

**Abstract.** Composite indices like the Gender Equality Index (GEI) are widely used to monitor gender disparities and guide evidence-based policy. However, their original design is often limited when applied to subnational contexts. Building on the GEI framework and the WeWorld Index Italia, this study proposes a composite indicator tailored to measure gender disparities across Italian regions. The methodology, based on a variation of the Mazziotta–Pareto Index, introduces a novel aggregation approach that penalizes uneven performances across domains. Indicators cover employment, economic resources, education, use of time, political participation, and health, reflecting multidimensional gender inequality. Using open regional data for 2024, the proposed Italian Gender Equality Index (IGEI) provides a comparable and robust measure across regions, highlighting both high-performing and lagging areas. The approach addresses compensatory limitations of traditional aggregation and offers a practical tool for regional monitoring and targeted interventions, benefiting from the fact that the IGEI is specifically tailored on the GEI framework.

## 1. Introduction

Gender inequality remains a persistent challenge across the globe, affecting economic opportunities, political representation, educational attainment, and health outcomes. Therefore, since effective policy design requires systematic and consistent measurement, in 2013 the European Institute for Gender Equality (EIGE) introduced the Gender Equality Index (GEI) as a comprehensive monitoring tool. This composite indicator is, nowadays, one of the leading international indices focusing on gender gaps, together with the Global Gender Gap Index (GGGI), first introduced in 2006 by the World Economic Forum. Although both indices aim to quantify and monitor gender inequalities, they differ substantially in methodology, scope, and normative orientation. The GGGI adopts a *gap-based* framework, measuring relative disparities between women and men independently of overall achievement levels. This approach emphasizes how equally resources and opportunities are distributed, irrespective of overall national performance, allowing it to highlight disparities even in low-income contexts. In contrast, the 2013 version of the GEI follows a *level-based* approach, jointly accounting for gender disparities and absolute achievement levels. Accordingly, the GGGI primarily serves as a tool to identify gender gaps across countries, whereas the 2013 GEI is explicitly designed to support evidence-based policymaking. Since its first publication, the Gender Equality Index has undergone several refinements, culminating in a substantial revision in the 2025 GEI Report. In this work, we adapt the GEI methodology to the Italian context, examining gender gaps within and between Italian regions. We introduce and justify structural modifications to the selection of indicators and the aggregation approach.

[1] Department of Enterprise Engineering “Mario Lucertini”, University of Rome Tor Vergata, Rome, Italy

The remainder of the paper is organized as follows. Section 2 outlines the conceptual framework of the Gender Equality Index, describing the domains of gender equality considered in this study and motivating their adaptation to the Italian regional context. Section 3 presents the methodological approach adopted for the construction of the Italian Gender Equality Index (IGEI), including the definition of the gender gap metric, the correcting coefficient, issues of time comparability, and the aggregation strategy across indicators, sub-domains, and domains. Section 4 describes the set of indicators used in the index, detailing their conceptual rationale, data sources, and construction for each domain. Section 5 presents the index results at the regional level, discusses territorial patterns with a few descriptive statistics, and compares the proposed index with alternative aggregation approaches and existing composite indicators. Finally, Section 6 concludes by summarizing the main contributions of the paper and discussing implications for policy analysis and future research. Technical proofs are reported in the Appendix, while descriptive tables are provided in the Annexes.

## 2. Conceptual framework

Since its first definition, the GEI is based on six different domains: *work*, *money*, *knowledge*, *time*, *power*, and *health*. The first domain, *work*, captures gender gaps in participation, segregation, and quality of employment within the labour market. The second domain, *money*, concerns gender gaps in financial resources and economic situation. The *knowledge* domain examines gender gaps in educational attainment and participation in education and training. The time domain addresses gender differences in the allocation of time, encompassing paid work, unpaid care activities, and participation in social, personal, and civic life (Eurofond, 2006). The fifth domain, *power*, focuses on gender gaps in representation and decision-making across political, social, and economic spheres, capturing women's and men's access to positions of power. The sixth domain, *health*, is central to gender equality, as health outcomes are closely linked to economic independence and intra-household bargaining power for women (Backhans et al., 2007). In the proposed Italian Gender Equality Index (IGEI), the domains considered are *work*, *money, knowledge*, *time*, *politics*, and *health*. These domains capture conceptual dimensions largely aligned with those of the 2025 GEI, albeit the GEI domains are more structured since they generally include a wider range of indicators. Following the structure of the GEI, each domain of the IGEI is further articulated into sub-domains, reflecting the multidimensional nature of gender inequality and allowing for a more granular assessment of disparities. Sub-domains group conceptually homogeneous indicators and are designed to distinguish between different mechanisms through which gender inequalities emerge.

**Table 1** – *IGEI domains and sub-domains*

| Domain | Work | | Money | | Knowledge | | Time | | Politics | | Health | |
|---|---|---|---|---|---|---|---|---|---|---|---|---|
| Subdomain | Participation | Entrepreneurship and quality of work | Labour earnings | Pension income | Attainment and participation | Segregation | Care activities | Social activities | National representation | Subnational representation | Status | Behaviour |

Within the *work* domain, sub-domains are defined to reflect, on the one hand, *participation* in the labour market and, on the other hand, structural features such as *entrepreneurship and quality of work*, which are closely linked to long-term economic autonomy. The *money* domain is structured to capture both *labour earnings* and *pension income*, acknowledging that gender inequalities in earnings tend to accumulate over the life course and manifest more strongly in older age. The *knowledge* domain is divided into sub-domains reflecting educational *attainment and participation* in learning activities, as well as gender *segregation* across fields of study, which plays a crucial role in shaping future labour market outcomes. The *time* domain distinguishes between *care activities* and *social activities*, reflecting the dual role played by unpaid care responsibilities and social engagement in shaping gendered time use. This distinction is particularly relevant in contexts where unequal care burdens constrain women's labour market participation and leisure opportunities. The *politics* domain focuses on political representation at different institutional levels and is structured to differentiate between *national* representation and *subnational representation*, recognizing that gender gaps in decision-making power may vary substantially across governance levels. Finally, the *health* domain is organized into sub-domains capturing health *status* and health-related *behaviour*.

This distinction reflects the understanding that gender differences in health outcomes are influenced both by biological factors and by social norms, lifestyles, and access to resources.

Violence, along with other intersecting inequalities, is not included among the main GEI domains due to significant limitations in data comparability, availability, and reliability across EU member states. As the GEI focuses on domains in which structural gender inequalities can be consistently measured over time, violence is addressed separately by EIGE through dedicated reports and indicators on gender-based violence. For the same reasons, we have decided not to include violence among the main IGEI domains.

## 3. Methodology

### *3.1. Existing GEI constructions*

To ensure comparability across populations with different demographic structures and sizes, most variables are expressed in relative terms. For example, employment rates are computed as the ratio of employed women and men to the working-age population. This normalization allows observed differences to be interpreted as gender gaps rather than reflections of population size or age composition. Here we introduce the notation used for such normalized variables:

$$x_i^k = \frac{X_i^k}{reference\ population_i^k}.$$

Here $k = w, m, a$ highlights whether the variable refers to women, men, or all the population, while $i = 1, \dots, 21$ denotes one of the NUTS2 statistical regions of Italy, namely the Italian regions with the Trentino-Alto Adige/Südtirol split into its two provinces. A similar notation is adopted in the GEI reports, where $i = 1, \dots, 27$ denotes the $i$-th European country. More specifically, the 2013 GEI methodology starts from the definition of a *gender gap metric*

$$1 - \gamma_i^{GEI,2013} = 1 - \left|1 - \frac{x_i^w}{x_i^a}\right|.$$

After specifying that the above normalized variables are assumed to be non-negative and positively correlated with the well-being of the corresponding reference population, we can claim that the gender gap metric yields values where 1 stands for complete gender equality, with any value below that indicating a proportional lack of gender equality, and with full gender inequality at 0. To standardize the indicators to have values between 1 and 100, initial 2013 GEI indicator was then calculated as:

$$1 + \left(1 - \gamma_i^{GEI,2013}\right) \cdot 99 \,.$$

However, a *correcting coefficient* was later introduced to ensure that obtaining a score close to 100 is the reflection of both low gender gaps and high levels of achievement. Such correcting coefficient has been defined as

$$\alpha_i^{GEI,2013} = \frac{x_i^a}{\max_i x_i^a} \,,$$

and the final indicator used in the 2013 GEI report was then

$$\Gamma_i^{\mathrm{GEI,2013}} = 1 + \alpha_i^{GEI,2013}\left(1 - \gamma_i^{GEI,2013}\right) \cdot 99 \,.$$

According to this formula, a first interpretation of the indicator $\Gamma_i^{GEI,2013}$ is that it reflects the *i*-th gender gap metric $1 - \gamma_i^{GEI,2013}$, but weighted by the relative level of average achievement $\alpha_i^{GEI,2013}$ and then scaled to a range from 1 to 100. The chosen standardization deliberately avoids 0 values to allow for later aggregation using geometric means. However, to better understand this indicator, let us assume $x_i^a \geq x_i^w > 0$ and notice that

$$\alpha_i^{GEI,2013}\left(1 - \gamma_i^{GEI,2013}\right) = \frac{x_i^w}{\max_i x_i^a} \,.$$

This clarifies how the GEI indicator $\Gamma_i^{GEI,2013}$ should be interpreted: it measures, after rescaling, the ratio of women's average well-being in the *i*-th country to that of the best-performing country. This approach highlights that the 2013 GEI's goal is not to measure local gender gaps, as the GGGI does using female-to-male ratios, but rather to encourage European countries to achieve, for all genders, the same level of well-being as that of the best-performing country.

Unfortunately, despite its revolutionary aspects, this approach does not penalize countries where men's well-being increases while women's remains constant. Some of these issues have been solved in the 2025 GEI version, where the GEI indicator is simply defined by

$$\Gamma_i^{\mathrm{GEI,2025}} = \frac{\min(x_i^w, x_i^m)}{\max\,(x_i^w, x_i^m)} \cdot 100 \,.$$

This new metric has been justified by the evolution of EU policy priorities in the last years, but it removes the level-based approach typical of the original GEI definition. It is therefore for such reasons that we want to suggest an alternative approach, explained in detail in this section.

*3.2. IGEI indicators*

Following the 2025 GEI Report, the starting point for the IGEI construction is to consider

$$\gamma_i = \frac{|x_i^w - x_i^m|}{\max\,(x_i^w, x_i^m)} = 1 - \frac{\min(x_i^w, x_i^m)}{\max\,(x_i^w, x_i^m)}\,.$$

This parameter is not only symmetric in the variables $x_i^w$ and $x_i^m$, ensuring an equal treatment of the two genders, but is also mathematically included in the bounded interval $[0,1]$ for any value of both $x_i^w$ and $x_i^m$. The *gender gap metric*

$$1 - \gamma_i = \frac{\min(x_i^w, x_i^m)}{\max\,(x_i^w, x_i^m)} \qquad (1)$$

is bounded to the interval $[0,1]$, is equal to 1 in the case of perfect gender balance $x_i^w = x_i^m$, and is equal to 0 whenever either $x_i^w = 0$ or $x_i^m = 0$.

We now proceed by defining our correcting coefficient. For the sake of brevity, in the following we will denote the best average with the notation

$$x^a = \max_i x_i^a\,.$$

The *correcting coefficient* we will use in the IGEI definition is then

$$\alpha_i = \frac{2x_i^a}{x^a + x_i^a}\,.$$

As before, this correcting coefficient is bounded to $[0,1]$ and is equal to 1 only when $x_i^a$ equals the maximum $x^a$. Relative to the 2013 GEI methodology, the proposed correcting coefficient is intentionally less severe, with the aim of preventing the corrective component from becoming overly dominant with respect to the core gender gap metric. In this new approach we are not comparing $x_i^a$ with the maximum average well-being $x^a$, but with the arithmetic mean

$$\frac{x^a + x_i^a}{2}\,.$$

We then define the *i*-th indicator of the Italian Gender Equality Index (IGEI) as

$$\Gamma_i = \alpha_i(1 - \gamma_i) \cdot 100\,.$$

The standardization has been modified to a 0-100 range to make the indicator more intuitive and easier to compute. Since geometric means are not later used for aggregation, this change does not raise any issues of mathematical consistency. The gender gap metric is aligned with the 2025 GEI definition, but the correcting coefficient is a less severe variation of the one used in the original 2013 GEI construction, avoiding poorly performing regions to be overshadowed by the top-performing ones. Therefore, this indicator follows a level-based approach and measures both gender gaps and average levels of achievement. Finally, differently from the 2013 GEI indicator, it depends on the gender gap within the $i$-th region. With classical analysis techniques indeed, it can be shown that, assuming to be in the $x_i^m \geq x_i^w > 0$ scenario, $\Gamma_i$ decreases as $x_i^m$ increases while $x_i^w$ remains constant. The $x_i^w \geq x_i^m > 0$ case follows by symmetry. However, rather than providing the detailed proof, we prefer to illustrate this scenario in a fictional case.

**Table 2** – *Values of* $\boldsymbol{\Gamma_i^{\mathrm{GEI,2013}}}$, $\boldsymbol{\Gamma_i^{GEI,2025}}$ *and* $\boldsymbol{\Gamma_i}$ *in a fictional case with six countries.*

| Country | $x_i^w$ | $x_i^m$ | $x_i^a$ | $\Gamma_i^{GEI,2013}$ | $\Gamma_i^{GEI,2025}$ | $\Gamma_i$ |
|---|---|---|---|---|---|---|
| A | 0.1 | 0.2 | 0.15 | 12.00 | 50.00 | 14.29 |
| B | 0.1 | 0.3 | 0.2 | 12.00 | 33.33 | 12.12 |
| C | 0.1 | 0.9 | 0.5 | 12.00 | 11.11 | 7.94 |
| D | 0.4 | 0.6 | 0.5 | 45.00 | 66.67 | 47.62 |
| E | 0.4 | 0.8 | 0.6 | 45.00 | 50.00 | 40.00 |
| F | 0.8 | 1.0 | 0.9 | 89.00 | 80.00 | 80.00 |

From this simple study we can notice that $\Gamma_i$ may be smaller or greater than $\Gamma_i^{GEI,2013}$, but its behaviour penalizes increasing gender gaps also in scenarios when the 2013 GEI index does not. Comparing the countries A and E, we can also show the IGEI methodology to be more affected by the average level of achievement than the 2025 GEI version. For the top-performing region F, the IGEI indicator $\Gamma_F$ is simply a rescaling of the ratio $x_F^w/x_F^m$. To provide a wider interpretation of our new index, however, let as assume $\Gamma_i$ and $\Gamma_j$ to be indicators associated with regions with same value for $x_i^a$. Assuming both $x_i^w \leq x_i^m$ and $x_j^w \leq x_j^m$, the identity

$$\frac{\Gamma_i}{\Gamma_j} = \frac{x_i^w/x_i^m}{x_j^w/x_j^m}$$

highlights how to read such ratio, namely it measures how much closer region $i$ is to gender parity compared with region $j$.

Finally, the last property we want to examine is time comparison. Technically, $\Gamma_i$ scores are comparable over time only if the maximum achievement $x^a$ remains constant over time. However, if we want to study the time series

$$\Gamma_{it} = \alpha_{it}(1-\gamma_{it})\cdot 100\,, \qquad t = t_1, \dots, t_n\,,$$

then we can modify the correcting coefficients

$$\alpha_{it} = \frac{2x_{it}^a}{x_{it}^a + x_t^a}$$

in such a way to make $\Gamma_{it}$ time comparable. More specifically, we can replace the maximum achievement at time $t$, here denoted by $x_t^a$, with

$$x^a = \max_{i,t} x_{it}^a\,.$$

By doing so, $\Gamma_{it}$ becomes time-comparable, namely changes over time of $\Gamma_{it}$ depends only on the $i$-th region and not on the other ones.

### 3.3. IGEI aggregation

We conclude this section by describing the aggregation methodology. In the 2025 GEI construction, each of the six domains of gender equality is further divided into two or three sub-domains. The final set consists of 27 indicators, all of them bounded between 0 and 100. These indicators have all positive polarity, namely each of them has score 0 in case of complete gender inequality and score 100 in case of full gender equality. To compute the overall index, a gradually compensatory aggregation method has been adopted. This means that the compensation allowed is higher within the aggregation at indicator level, while it becomes gradually less compensatory within subdomain and domain levels. More specifically, it consists of an arithmetic mean with equal weights at indicator level, a geometric mean with equal weights at sub-domain level, and a geometric mean with experts' weights at domain level. The six 2025 GEI domain weights were obtained through an online survey conducted in the last quarter of 2024, which gathered responses from 125 EIGE stakeholders. Weights were derived using the analytic hierarchy process (AHP), a structured method based on expert judgement and matrix diagonalisation.

The IGEI consists of 22 indicators distributed across six domains and twelve sub-domains, as described below. The *money* domain includes two indicators, each corresponding to a distinct sub-domain. The *work* and *politics* domains each comprise three indicators distributed across two sub-domains. The *knowledge* and *time* domains each aggregate four indicators, with two indicators per sub-domain. Finally, the *health* domain includes six indicators, with three indicators assigned to each sub-domain. Indicators of the same sub-domain are aggregated with an arithmetic mean, but for aggregating different subdomains indicators we choose to make use of the penalized arithmetic mean defined in the Appendix. More explicitly, let $\Gamma_{ids\,1}, \dots, \Gamma_{idsj_m}$ be the complete set of indicators belonging to the same domain $d$ and sub-domain $s$ for a chosen region $i$. The corresponding sub-domain indicator is then defined as

$$I_{ids} = \frac{\Gamma_{idsj_1} + \cdots + \Gamma_{idsj_m}}{m} \ ,$$

with $m = m_s$ the number of indicators included in the sub-domain $s$. The next step is performed by using a more penalizing aggregation methodology, namely the domain indicator $I_{id}$ is defined as

$$I_{id} = \frac{I_{id\,1} + \cdots + I_{ids_n}}{n} - \frac{Var\left(I_{ids_1}, \dots, I_{ids_n}\right)}{2(\max I_{ids} - \min I_{ids})} \ ,$$

where $Var$ denotes the variance, $\max I_{ids}$ and $\min I_{ids}$ denote respectively the maximum and minimum values of all indicators within the same sub-domain $s$, and $n = n_s$ is the number of indicators included in the sub-domain $s$. To compute the final indicator $I_i$, the same aggregation methodology is applied. Therefore, we can finally define the *Italian Gender Equality Index (IGEI)* for the *i*-th region as:

$$I_i = \frac{I_{i1} + \cdots + I_{i6}}{6} - \frac{Var(I_{i1}, \dots, I_{i6})}{2(\max I_{id} - \min I_{id})} \, .$$

With this formula we conclude the general overview on the mathematical approach used in the IGEI definition. The next section will focus on describing the gender indicators used for this construction.

## 4. Indicators

This section presents the set of 22 indicators included in the IGEI domains. The selection of indicators draws on both the Gender Equality Index and the WeWorld Index Italia. The latter is an annual report developed by the WeWorld-GVC Foundation, an Italian NGO, which measures the inclusion and quality of life of women and children across Italy's regions. The foundation also produces the WeWorld Index, a similar tool with a global scope.

The domain of *work* compares men's and women's inclusion in the labour market. This domain includes two conceptual sub-domains to be measured: *participation*, and *entrepreneurship and quality of work*. Participation is measured by the *employment rate* ($\Gamma_1$), with such rate referring to the 20-64 age interval and to 2024 Istat data. The second sub-domain is combined with two gender indicators: *Entrepreneurial density* ($\Gamma_2$) and *involuntary part-time employment* ($\Gamma_3$).

**Table 3** – *IGEI indicators*

| Domain | Subdomain | No. | Indicator |
|---|---|---|---|
| Work | Participation | 1 | Employment rate |
| | Entrepreneurship and quality of work | 2 | Entrepreneurial density |
| | | 3 | Involuntary part-time employment |
| Money | Labour earnings | 4 | Average annual gross wage of employees |
| | Pension income | 5 | Average annual per-capita pension income |
| Knowledge | Attainment and participation | 6 | Tertiary graduates and other tertiary qualifications |
| | | 7 | Participation in lifelong learning |
| | Segregation | 8 | Graduates in female-dominated fields |
| | | 9 | Graduates in male-dominated fields |
| Time | Care activities | 10 | Ratio between employment rates of women with preschool children and women without children |
| | | 11 | Authorized places in early childhood education and care services |
| | Social activities | 12 | Social participation |
| | | 13 | Volunteering activities |
| Politics | National representation | 14 | Women and political representation in Parliament |
| | Subnational representation | 15 | Women and political representation at the local level |
| | | 16 | Women in municipal administrators |
| Health | Status | 17 | Life expectancy at birth |
| | | 18 | Healthy life expectancy at birth |
| | | 19 | Mental health |
| | Behaviour | 20 | Smoking |
| | | 21 | Alcohol |
| | | 22 | Adequate nutrition |

The computation of the indicator $\Gamma_1$ follows the methodology described in the previous section, since employment rates are provided on a region level for women, men, and for the total population. Entrepreneurial density $\Gamma_2$ measures the gender inequalities in the different entrepreneurial opportunities. More in general, a high entrepreneurial density indicates that entrepreneurship is relatively widespread and that individuals are more likely to engage in business activity, even though a too high density may be a symptom of a fragmented economic structure. The entrepreneurial density is defined as the number of registered enterprises relative to the labour force (*popolazione attiva*) aged 15-64. Female labour force is used as a reference population for the number of female-owned businesses (*imprese femminili*), while the male labour force is used as a reference population for all the other businesses. The notion of women-owned business (*impresa femminile*) follows the 2009's algorithm of Unioncamere – Infocamere. The construction of the $\Gamma_2$ indicator, based on 2024 data, follows the general methodology, but no correcting coefficient has been used. A different approach applies to the indicator $\Gamma_3$, because in this case the studied variables $x_i^w, x_i^m$ and $x_i^a$ have all negative polarity. However, since these variables are rates belonging to the 0-1 interval, an inversion of polarity can be easily achieved. The transformed variables are then

$$1 - x_i^k\,, \qquad k = w, m, a\,.$$

After performing this reversion, the methodology described in formula (1) is applied.

The domain of *money* focuses on financial equality aspects such as the gender pay gap. Its two indicators are the *average annual gross wage of employees* ($\Gamma_4$) and the *average annual per-capita amount pension income* ($\Gamma_5$). In this case no modification is performed to the indicator definition described in the previous section. Data are collected by Istat sources and refer to 2023's wages and pension incomes.

The domain of *knowledge* measures both gender differences in educational attainment and gender segregation in education. This combination leads to the two sub-domains of *attainment and participation* and *segregation*. The first sub-domain includes an indicator focusing on *tertiary graduates and other tertiary qualifications* ($\Gamma_6$) and an indicator focusing on *participation in lifelong learning* ($\Gamma_7$). The $\Gamma_6$ indicator's reference population is given by women and men of age 25-34, while the $\Gamma_7$ indicator refers to people of age 25-64 who attended some education or training activity in the four weeks preceding the interview. Both these indicators have been computed using 2024 Istat data and with the standard IGEI methodology. The second sub-domain corresponds to two indicators measuring, respectively, *graduates in female-dominated fields* ($\Gamma_8$) and *graduates in male-dominated fields* ($\Gamma_9$). Indicator $\Gamma_8$ has been computed with MUR open data, and concerns people graduated in 2024 in education and training, art and design, literary studies and humanities, and in the medical, health and pharmaceutical field. The choice of these "female-dominated" fields follows a similar indicator used in the GEI construction. In this case, the variable for the correcting coefficient is not the total number of graduates in female-dominated fields, as this does not capture gender inequalities. Instead, following GEI reports and expert studies, the rate of tertiary education graduates has been adopted, since higher levels of participation in tertiary education increase the potential for gender-based segregation in fields of study (Hakim, 1996; Charles and Bradley, 2002). Apart from this difference, the definition of this indicator still follows the approach described in the previous section. The same methodology applies to the $\Gamma_9$ indicator, where the notion of "male-dominated" fields refers to similar choices made in the construction of the 2025 GEI.

The domain of *time* attempts to capture the gendered nature of the allocation of the time spent between childcare and domestic tasks, and how this gap affects work-life balance, as well as social and volunteering activities. This domain is therefore composed of two sub-domains: *care activities* and *social activities*. The first sub-domain is composed of two indicators, namely *ratio between employment rates of women with preschool children and women without children* ($\Gamma_{10}$), and *authorized places in early childhood education and care services* ($\Gamma_{11}$). The first indicator focuses on women aged 25-49 and with at least one daughter or son aged 0-5 and refers to 2024 Istat data. The indicator's construction follows the general IGEI approach, even though in this case the gender metric reads

$$1 - \gamma_i = \min\left(x_i, x_i^{-1}\right), \tag{2}$$

with $x_i$ the aforementioned ratio. This expression still follows the theoretical discussion of the previous section: by writing

$$x_i = \frac{x_i^{w,children}}{x_i^{w,no\ children}},$$

we can notice how the previous metric can be written as

$$1 - \gamma_i = \frac{\min\left(x_i^{w,children}, x_i^{w,no\ children}\right)}{\max\left(x_i^{w,children}, x_i^{w,no\ children}\right)},$$

which leads us to equation (2). As a correcting coefficient, in this case we adopted the ratio of employed females already used for the $\Gamma_1$ indicator. The $\Gamma_{11}$ indicator, focusing on education and care services for children aged 0-2, is quite different from the other indicators as it does not explicitly mention any gender gap. However, inspired by the works on the WeWorld Index Italia, we decided to include this indicator as it is often correlated with gender gaps in time spent for childcare and domestic tasks. Due to its different nature, no correcting coefficient is applied, and its definition is simply

$$\Gamma_{i11} = \min(1, x_i) \cdot 100\,. \qquad (3)$$

The corresponding Istat data refers to 31st December 2022. With respect to the social activities sub-domain the considered indicators are *social participation* ($\Gamma_{12}$) and *volunteering activities* ($\Gamma_{13}$). In particular, the volunteering and charity area has been identified as important in gender terms because it can be constructed as an extension of women's responsibility for caring activities (Neysmith and Reitsma-Street, 2000). These indicators refer to people at least 14 years old who participated in some social or volunteering activity at least once in the 12 months preceding the interview. In both cases, we applied the general IGEI construction to 2024 Istat data.

The domain of *politics* is measured by three gender indicators that examine representation in parliaments, regional assemblies, and municipalities: *women and political representation in parliament* ($\Gamma_{14}$), *women and political representation at the local level* ($\Gamma_{15}$), and *women in municipal administration* ($\Gamma_{16}$), with the last two aggregated in the *subnational representation* domain. For such indicators, the correcting coefficient was not applied because these variables represent shares. For example, assume $x_i^w$ and $x_i^m$ to represent the share of parliamentary seats held, respectively, by women and men of a given region. Since $x_i^w + x_i^m = 1$, the IGEI indicator can be written as

$$\Gamma_i = \min\left(\frac{x_i^w}{1 - x_i^w}, \frac{1 - x_i^w}{x_i^w}\right) \cdot 100\,. \qquad (4)$$

The same argument applies for the indicator measuring the political representation at the local level. All these indicators are based on 2024 Istat data.

The final core domain examines issues related to gender and health. It includes two sub-domains: *status* and *behaviour*. For the health status sub-domain, the selected gender indicators are *life expectancy at birth* ($\Gamma_{17}$), *healthy life expectancy at birth* ($\Gamma_{18}$) and *mental health* ($\Gamma_{19}$). For what concerns the first two indicators, it is important to point how that, despite women tend to outlive men around the world, healthy life expectancy reflects not just biological differences, but also social and economic conditions that affect well-being. Following the GEI approach, we decided to include both these indicators. All indicators of the status sub-domain are computed by following the standard IGEI methodology. The last sub-domain, behaviour, includes three indicators respectively measuring *smoking* ($\Gamma_{20}$), *alcohol* ($\Gamma_{21}$), and *adequate nutrition* ($\Gamma_{22}$). These indicators follow the standard IGEI methodology, but the variables used for the first two indicators have been inverted as already shown with the $\Gamma_2$ indicator. All the variables are based on 2024 Istat data.

This section has provided a descriptive analysis of the gender indicators considered in the Italian Gender Equality Index. For each critical area of gender equality, the gender gaps in each selected gender indicator depict the situation across Italian regions. It is by focusing on each of these critical areas of that policymaking can effectively contribute to progress in making gender equality a reality. We conclude this section by providing a metadata table with details about the gender indicators.

**Table 4** – *IGEI processing of variables*

| Indicator | Update | Reversion | Metric | Correction |
|---|---|---|---|---|
| Employment rate | 2024 | + | (1) | Original variable |
| Entrepreneurial density | 2024 | + | (1) | Not used |
| Involuntary part-time employment | 2024 | - | (1) | Original variable |
| Average annual gross wage of employees | 2023 | + | (1) | Original variable |
| Average annual per-capita pension income | 2023 | + | (1) | Original variable |
| Tertiary graduates and other tertiary qualifications | 2024 | + | (1) | Original variable |
| Participation in lifelong learning | 2024 | + | (1) | Original variable |
| Graduates in female-dominated fields | 2024 | + | (1) | Tertiary graduates and other tertiary qualifications |
| Graduates in male-dominated fields | 2024 | + | (1) | Tertiary graduates and other tertiary qualifications |
| Ratio between employment rates of women with preschool children and women without children | 2024 | + | (2) | Employment rate of women |
| Authorized places in early childhood education and care services | 2022 | + | (3) | Not used |
| Social participation | 2024 | + | (1) | Original variable |
| Volunteering activities | 2024 | + | (1) | Original variable |
| Women and political representation in Parliament | 2022 | + | (4) | Not used |
| Women and political representation at the local level | 2024 | + | (4) | Not used |
| Women in municipal administrators | 2024 | + | (4) | Not used |
| Life expectancy at birth | 2024 | + | (1) | Original variable |
| Healthy life expectancy at birth | 2024 | + | (1) | Original variable |
| Mental health | 2024 | + | (1) | Original variable |
| Smoking | 2024 | - | (1) | Original variable |
| Alcohol | 2024 | - | (1) | Original variable |
| Adequate nutrition | 2024 | + | (1) | Original variable |

NB: The number in the fourth column corresponds to the equation describing the gender gap metric.

## 5. Index results

### 5.1. IGEI scores and statistics

The Italian Gender Equality Index attempts to provide a comprehensive measure of the gender gaps across the Italian regions. Building upon this work, this section now focuses on the scores of the Gender Equality Index in each region and in Italy overall to offer a detailed assessment of where Italian regions stand in achieving gender equality. We will also compare the constructed index with others already existing and highlight how different aggregation methodologies can change the results.

**Table 5** – *IGEI regional scores*

| Region | IGEI | Work | Money | Knowledge | Time | Politics | Health |
|---|---|---|---|---|---|---|---|
| Emilia-Romagna | 66.76 | 68.25 | 67.66 | 59.02 | 56.44 | 73.55 | 85.00 |
| Umbria | 64.39 | 67.98 | 63.69 | 65.50 | 59.01 | 57.98 | 79.17 |
| Friuli-Venezia Giulia | 63.37 | 69.20 | 64.34 | 52.74 | 54.20 | 66.06 | 84.01 |
| Toscana | 63.33 | 68.11 | 65.75 | 58.58 | 56.22 | 57.29 | 83.85 |
| Veneto | 62.89 | 67.63 | 63.52 | 58.06 | 54.02 | 60.74 | 81.99 |
| Provincia Autonoma di Trento | 62.17 | 67.74 | 60.64 | 49.41 | 60.99 | 58.98 | 85.42 |
| Lazio | 60.77 | 65.21 | 70.01 | 59.95 | 51.22 | 45.47 | 85.44 |
| Piemonte | 60.71 | 68.85 | 67.91 | 53.15 | 55.27 | 46.73 | 84.91 |
| Trentino-Alto Adige/Südtirol | 60.69 | 69.28 | 62.41 | 43.66 | 55.26 | 61.85 | 82.75 |
| Valle d'Aosta/Vallée d'Aoste | 60.60 | 71.97 | 62.62 | 37.94 | 61.27 | 60.84 | 81.06 |
| Lombardia | 59.58 | 67.08 | 70.82 | 53.40 | 55.60 | 42.96 | 79.75 |
| Marche | 58.80 | 69.34 | 63.44 | 53.58 | 54.38 | 41.25 | 83.46 |
| **Italia** | **58.20** | **64.92** | **66.00** | **52.73** | **49.24** | **46.44** | **82.66** |
| Provincia Autonoma di Bolzano/Bozen | 57.61 | 70.80 | 64.10 | 38.03 | 50.44 | 56.28 | 79.23 |
| Abruzzo | 57.01 | 66.10 | 58.97 | 58.67 | 45.04 | 44.14 | 82.36 |
| Liguria | 57.01 | 66.05 | 62.87 | 51.55 | 52.08 | 37.48 | 86.03 |
| Sardegna | 55.83 | 62.81 | 61.15 | 53.32 | 48.49 | 40.55 | 80.68 |
| Calabria | 52.90 | 53.09 | 57.67 | 50.83 | 36.78 | 47.43 | 85.41 |
| Molise | 52.63 | 63.99 | 58.34 | 47.18 | 50.74 | 29.33 | 80.70 |
| Sicilia | 52.33 | 53.91 | 57.77 | 46.20 | 33.83 | 52.09 | 83.93 |
| Puglia | 49.58 | 54.74 | 56.75 | 44.60 | 43.69 | 32.78 | 78.26 |
| Basilicata | 49.10 | 59.42 | 58.07 | 45.79 | 45.70 | 21.48 | 80.33 |
| Campania | 47.49 | 52.03 | 57.32 | 44.84 | 32.08 | 33.50 | 82.25 |

The map in Figure 1 illustrates the division of Italian regions into macro-areas based on their IGEI scores. The highest-performing regions are located in the north-east and central-north, with Emilia-Romagna, Umbria, and Friuli-Venezia Giulia emerging as the top performers. The lowest-performing regions are in southern Italy, with Puglia, Basilicata, and Campania ranking at the bottom.

**Figure 1** – *IGEI national map, 2024*

Across regions, *health* consistently records the highest scores, exhibiting limited territorial variation, while *politics* and *time* emerge as the weakest and most heterogeneous domains. Following the *knowledge* domain, economic dimensions *work* and *money* display intermediate but persistent gaps. Notably, the two autonomous provinces display domain-specific asymmetries.

**Table 6** – *Descriptive statistics for IGEI domains*

| Index | mean | sd | cv | min | p25 | p50 | p75 | max |
|---|---|---|---|---|---|---|---|---|
| IGEI | **57.85** | **5.32** | **0.09** | **47.49** | **52.90** | **58.80** | **62.17** | **66.76** |
| Work | 64.49 | 5.99 | 0.09 | 52.03 | 62.81 | 67.08 | 68.25 | 71.97 |
| Money | 62.54 | 4.15 | 0.07 | 56.75 | 58.34 | 62.87 | 64.34 | 70.82 |
| Knowledge | 51.54 | 6.99 | 0.14 | 37.94 | 46.20 | 52.74 | 58.06 | 65.50 |
| Time | 50.36 | 8.07 | 0.16 | 32.08 | 45.70 | 52.08 | 55.60 | 61.27 |
| Politics | 47.95 | 12.82 | 0.27 | 21.48 | 40.55 | 46.73 | 57.98 | 73.55 |
| Health | 82.53 | 2.37 | 0.03 | 78.26 | 80.68 | 82.36 | 84.91 | 86.03 |

NB: The statistics are computed selecting only the 21 NUTS2 statistical Italian regions.

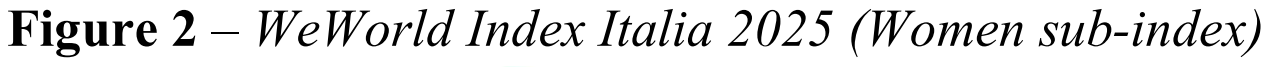

**Figure 2** – *WeWorld Index Italia 2025 (Women sub-index)*

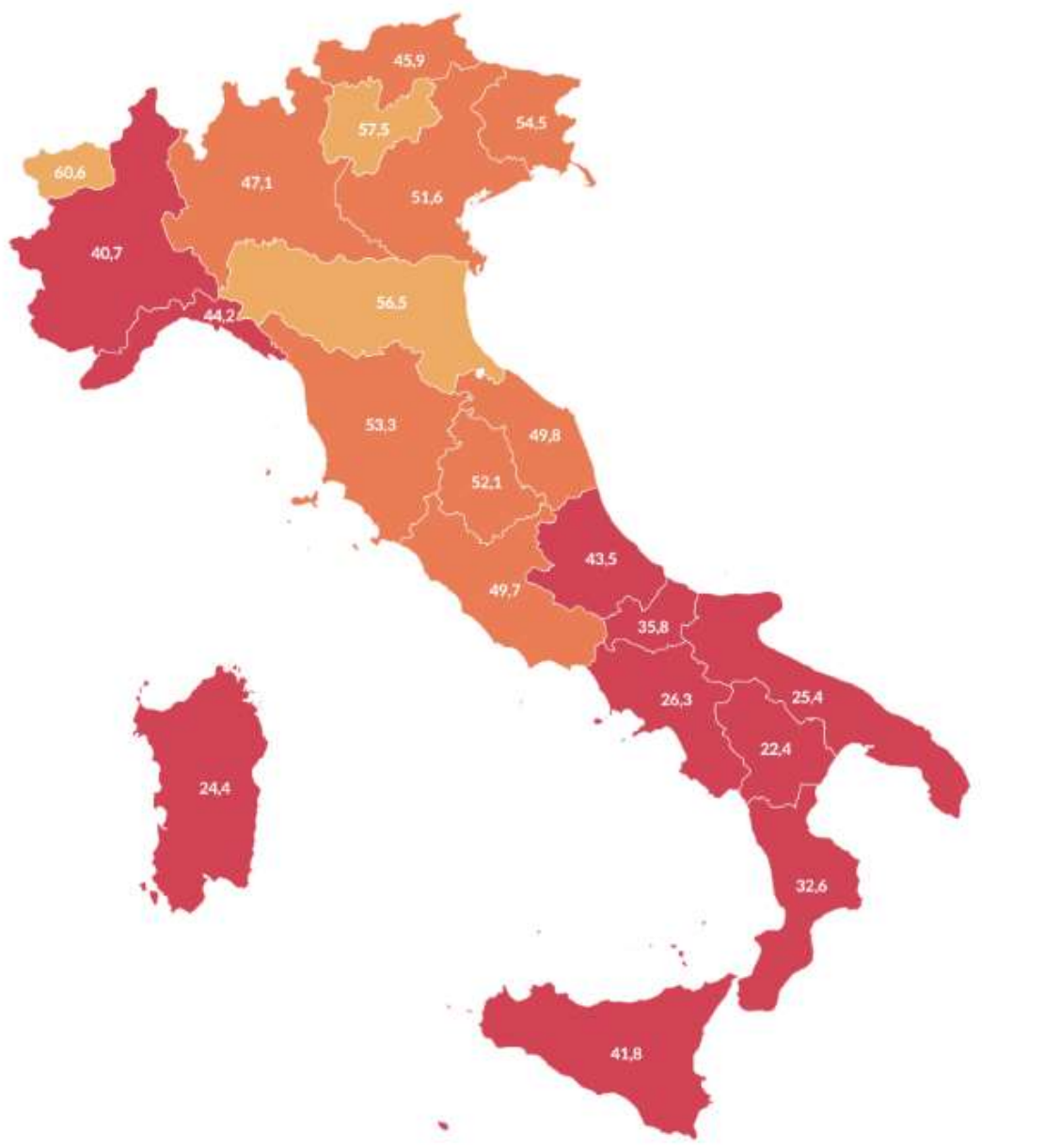

As shown in Figure 2, our results are partially described by the findings of the WeWorld Index Italia 2025. This composite indicator includes a sub-indicator specifically measuring the social and economic conditions of women. Here too, a notable south-to-north gradient in the index is observed.

### *5.2. Alternative indices*

To better assess our results, we now make a few considerations about the aggregation strategy. As described in the methodology section, our approach relies on an increasingly penalising aggregation, namely we use an arithmetic mean at the variable level, and the penalized arithmetic mean defined in the Appendix at both the sub-domain and domain levels. As this penalized arithmetic mean construction has been inspired by the *Adjusted Mazziotta-Pareto Index* (AMPI), we will apply the AMPI strategy to the IGEI 22 gender indicators and compare the different rankings. Before proceeding, we briefly recall the AMPI procedure (Mazziotta and Pareto 2016), a variation of the original Mazziotta-Pareto Index (De Muro et al. 2011).

Consider a matrix $X = (x_{ij})$ with $n$ rows (statistical units) and $m$ columns (statistical indicators). All the indicators are assumed to have positive *polarity*, namely they are assumed to be positively correlated with the phenomenon to be measured. Let us compute the standardized matrix $R = (r_{ij})$ as follows:

$$r_{ij} = \frac{x_{ij} - g_{min,j}}{g_{max,j} - g_{min,j}} 60 + 70 \, ,$$

where $g_{min,j}$ and $g_{max,j}$ are the *goalposts* for the $j$-th indicator. The *goalposts* $g_{min,j}$ and $g_{max,j}$ define the lower and upper reference bounds used to normalize each indicator without using a strict min-max standardization, but still ensuring that, in most common cases, the new standardized values are included between 70 and 130 like a normal distribution with mean 100 and standard deviation 10. Explicitly, given a reference value $R_j$, the goalposts for the $j$-th column $x_j$ are defined as

$$\begin{cases} g_{min,j} = R_j - \Delta_j \\ g_{max,j} = R_j + \Delta_j \end{cases},$$

where $\Delta_j = \max x_j - \min x_j$. After the standardization step, the AMPI for the $i$-th statistical unit is defined as

$$AMPI = M_i - S_i \cdot cv_i \,, \tag{5}$$

where $M_i$, $S_i$, and $cv_i = S_i / M_i$ are respectively the mean, the standard deviation, and the coefficient of variation of the $i$-th standardized unit $r_i$, namely the $i$-th row of $R$. In our case, the reference value for each indicator is the corresponding Italy score. By doing so, regions with an indicator value over (below) the national average will have a standardized value larger (smaller) than 100.

**Table 7** – *IGEI variations scores and rankings*

| Regione | IGEI | IGEI Rank | AMPI | AMPI Rank | AMI | AMI Rank | GMI | GMI Rank |
|---|---|---|---|---|---|---|---|---|
| Emilia-Romagna | 66.76 | 1 | 112.64 | 1 | 70.51 | 1 | 69.14 | 1 |
| Umbria | 64.39 | 2 | 106.31 | 3 | 67.66 | 3 | 66.76 | 2 |
| Friuli-Venezia Giulia | 63.37 | 3 | 105.84 | 4 | 67.92 | 2 | 65.61 | 3 |
| Toscana | 63.33 | 4 | 106.57 | 2 | 66.30 | 4 | 65.36 | 4 |
| Veneto | 62.89 | 5 | 102.21 | 10 | 66.01 | 6 | 64.86 | 5 |
| Provincia Autonoma di Trento | 62.17 | 6 | 102.61 | 9 | 65.80 | 7 | 64.07 | 6 |
| Lazio | 60.77 | 7 | 104.70 | 6 | 64.32 | 9 | 62.61 | 8 |
| Piemonte | 60.71 | 8 | 105.17 | 5 | 64.60 | 8 | 62.80 | 7 |
| Trentino-Alto Adige/Südtirol | 60.69 | - | 101.72 | - | 65.49 | - | 62.67 | - |
| Valle d'Aosta/Vallée d'Aoste | 60.60 | 9 | 102.91 | 8 | 66.04 | 5 | 61.91 | 9 |
| Lombardia | 59.58 | 10 | 103.27 | 7 | 63.06 | 11 | 61.33 | 10 |
| Marche | 58.80 | 11 | 101.27 | 11 | 62.47 | 12 | 60.66 | 11 |
| **Italia** | 58.20 | - | 100.00 | - | 61.60 | - | 60.06 | - |
| Provincia Autonoma di Bolzano/Bozen | 57.61 | 12 | 98.06 | 13 | 63.24 | 10 | 59.17 | 12 |
| Abruzzo | 57.01 | 13 | 94.11 | 15 | 60.72 | 13 | 58.77 | 13 |
| Liguria | 57.01 | 14 | 98.12 | 12 | 60.61 | 14 | 58.33 | 14 |
| Sardegna | 55.83 | 15 | 94.41 | 14 | 59.44 | 15 | 57.59 | 15 |
| Calabria | 52.90 | 16 | 87.46 | 17 | 57.70 | 16 | 54.67 | 16 |
| Molise | 52.63 | 17 | 89.33 | 16 | 56.40 | 18 | 53.59 | 18 |
| Sicilia | 52.33 | 18 | 87.12 | 18 | 56.81 | 17 | 54.09 | 17 |
| Puglia | 49.58 | 19 | 83.59 | 20 | 53.45 | 20 | 51.00 | 19 |
| Basilicata | 49.10 | 20 | 85.56 | 19 | 53.63 | 19 | 49.36 | 20 |
| Campania | 47.49 | 21 | 80.55 | 21 | 52.41 | 21 | 49.01 | 21 |

After the standardization step, the indicators are aggregated first by using an arithmetic mean, and then by using formula (5) twice like we did with the penalized arithmetic mean in the IGEI construction. In the comparison table, we also provide scores and rankings of two other IGEI variations. In the first case we replace the penalized arithmetic mean with the classical arithmetic mean to construct an arithmetic mean index (AMI). In the latter, we replace the penalized arithmetic mean with the geometric mean to obtain some geometric mean index (GMI). In all these IGEI variations, Emilia-Romagna and Campania always occupy the first and the last positions in the final ranking.

We conclude this section with one last comparison of different aggregation approaches. As described in the methodology section, the 2025 Gender Equality Index is built on 27 indicators organized into 13 sub-domains and 6 domains. These 27 indicators are aggregated using an arithmetic mean, then the sub-domain indicators are aggregated with a geometric mean using equal weights, and finally the domain indicators are aggregated with a geometric mean using unequal weights determined through an AHP process. We now attempt to replace the geometric mean with our penalized arithmetic mean to construct a new ranking of the 2025 GEI. Since the AHP domain weights were published in the 2025 GEI report only in rounded form (two decimal places), we estimated the original weights using a reverse-engineering computational approach based on exhaustive combinatorial analysis. The core idea was to apply a logarithmic transformation to the weighted geometric mean of the six domains to create a linear equation for each EU member state, automatically solve all the linear systems given by all possible combinations of 5 countries, and identify the weight values that best approximate the published 2025 GEI index values for all EU member states. Further details are available in the public GitHub repository lpanebianco/GEI2025weightsAnalysis.

**Table 8** – *Approximations of the 2025 GEI domain weights*

| Work | Money | Knowledge | Time | Power | Health |
|---|---|---|---|---|---|
| 0.16 | 0.18 | 0.13 | 0.19 | 0.18 | 0.16 |

NB: The weights, to 15 decimal places, used in this paper in the calculation of the GEI variations are the following: work 0,162511466842102; money 0,181002399159846; knowledge 0,12726846491938; time 0,187055022839716; power 0,18376386124; health 0,158398784988964

Finally, we compare the 2025 GEI scores and ranking with other indexes based on the same indicators but different aggregation approaches at both the sub-domain and the domain levels. In the first case we replace the geometric means (with equal and unequal weights) with the penalized arithmetic means (with equal and unequal weights) to define the Penalized Gender Equality Index (PGEI). In the second scenario, we perform a similar variation but applying the classical arithmetic mean with same weights, and the corresponding index is referred to as Arithmetic Gender Equality Index (AGEI).

**Table 9** – *GEI variations scores and rankings*

| Member State | GEI | GEI Rank | PGEI | PGEI Rank | AGEI | AGEI Rank |
|---|---|---|---|---|---|---|
| SE | 73.7 | 1 | 71.6 | 1 | 75.5 | 1 |
| FR | 73.4 | 2 | 71.5 | 2 | 74.7 | 2 |
| DK | 71.8 | 3 | 69.3 | 3 | 74.0 | 3 |
| ES | 70.9 | 4 | 69.0 | 4 | 72.1 | 4 |
| NL | 69.5 | 5 | 67.4 | 5 | 71.1 | 7 |
| IE | 69.0 | 6 | 67.0 | 6 | 71.2 | 6 |
| BE | 68.5 | 7 | 66.5 | 7 | 71.2 | 5 |
| FI | 68.3 | 8 | 66.2 | 8 | 70.7 | 8 |
| LU | 63.9 | 9 | 62.3 | 9 | 67.2 | 9 |
| PT | 63.4 | 10 | 61.8 | 10 | 66.9 | 10 |
| **EU-27** | 63.4 | - | 61.7 | - | 66.1 | - |
| DE | 63.2 | 11 | 61.4 | 11 | 66.5 | 12 |
| IT | 61.9 | 12 | 59.8 | 14 | 63.8 | 18 |
| AT | 61.2 | 13 | 60.1 | 13 | 64.7 | 15 |
| LT | 60.9 | 14 | 59.5 | 15 | 64.8 | 14 |
| EE | 59.4 | 15 | 60.4 | 12 | 66.7 | 11 |
| MT | 58.9 | 16 | 58.6 | 18 | 62.9 | 21 |
| BG | 58.1 | 17 | 57.8 | 20 | 63.4 | 19 |
| SI | 58.0 | 18 | 58.8 | 17 | 65.2 | 13 |
| PL | 57.8 | 19 | 59.0 | 16 | 64.6 | 16 |
| SK | 57.2 | 20 | 57.2 | 21 | 63.3 | 20 |
| HR | 57.1 | 21 | 58.1 | 19 | 63.8 | 17 |
| EL | 57.0 | 22 | 56.6 | 22 | 61.8 | 22 |
| RO | 57.0 | 23 | 56.3 | 23 | 61.3 | 24 |
| LV | 56.7 | 24 | 55.5 | 24 | 61.2 | 25 |
| CZ | 53.2 | 25 | 53.8 | 26 | 59.9 | 26 |
| HU | 51.6 | 26 | 55.4 | 25 | 61.7 | 23 |
| CY | 47.6 | 27 | 50.1 | 27 | 56.0 | 27 |

Similarly to the previous case, the four highest-ranked countries (Sweden, France, Denmark, and Spain) and the lowest-ranked country (Cyprus) maintain the same ranking across all possible scenarios. AGEI scores are consistently higher than the corresponding GEI and PGEI scores, whereas PGEI scores exceed GEI scores in fewer than one third of the cases. This pattern highlights that, for indicators whose values are already standardized on a 0–100 scale, such as the GEI, the penalized arithmetic mean represents a valid and more stringent aggregation method, as it rewards countries that achieve high and balanced performance across all domains.

## 6. Conclusions

This work has presented the construction of a gender equality index tailored to Italian regions, offering a detailed and updated picture of gender inequalities across the country. The index reveals marked territorial disparities, with north-eastern and central-northern regions generally performing better, and southern regions continuing to lag behind. These results are consistent with the latest findings of the European Institute for Gender Equality (EIGE), which placed Italy below the EU average in the 2025 Gender Equality Index, with a score of 61.9 out of 100 (compared to the EU average of 63.4). In particular, the lower regional scores in Italy reflect the national score's relative position below the EU average, confirming persistent gender gaps especially in certain domains. Among the six IGEI domains analysed, as shown in the table below, the greatest regional disparities were observed in the domains of *politics* (standard deviation: 12.82, coefficient of variation: 0.27), *time* (8.07, 0.16), and *knowledge* (6.99, 0.14). These findings point to the need for targeted interventions at the regional level, focusing on improving political representation, balancing time use between work and care responsibilities, and enhancing access to education and skills development to reduce gender inequalities in the most critical areas.

## 7. Acknowledgments

The author would like to thank Professor Mazziotta for his inspiring course on composite indicators and for his valuable suggestions that contributed to the development of this work.

## 8. Appendix

In this appendix we prove a few technical mathematical results which introduce and motivate the aggregation method used in this work. Such a construction has been inspired by the works of Mazziotta and Pareto and the following theorem.

*Theorem (Cartwright and Field, 1978):* Suppose that $x_k \in [a,b]$ and $p_k \geq 0$ for $k = 1, \dots, n$, where $a > 0$, and suppose that $\sum_k p_k = 1$. Then, writing $\overline{x} = \sum_k p_k x_k$, we have

$$\frac{1}{2b}\sum_{k=1}^{n} p_k (x_k - \overline{x})^2 \leq \overline{x} - \prod_{k=1}^{n} x_k^{p_k} \leq \frac{1}{2a}\sum_{k=1}^{n} p_k (x_k - \overline{x})^2 \,. \qquad (A1)$$

This theorem represents a refinement of the classical arithmetic mean-geometric mean inequality. In a more concise way, if we denote by $\overline{x}_g$ the weighted geometric mean and by $var(x)$ the weighted variance appearing on both the left and right sides, we can write $(A1)$ as

$$\frac{var(x)}{2a} \leq \overline{x} - \overline{x}_g \leq \frac{var(x)}{2b} \,.$$

Furthermore, in the proof of this theorem it is implicitly shown that the constants $1/2a$ and $1/2b$ are optimal, namely

$$\frac{1}{2a} = \sup_{a \leq x_k} \frac{\overline{x} - \overline{x}_g}{var(x)} \,, \qquad \frac{1}{2b} = \inf_{x_k \leq b} \frac{\overline{x} - \overline{x}_g}{var(x)} \,.$$

We now use the Cartwright-Field Theorem to prove the following result.

*Theorem:* Let $x_k$ be a finite sequence of $n$ elements. If $r = \max_k x_k - \min_k x_k$ is not null, then

$$\min_k x_k \leq \overline{x} \pm \frac{1}{2r}\sum_{k=1}^{n} p_k (x_k - \overline{x})^2 \leq \max_k x_k \,. \qquad (A2)$$

*Proof:* Given a finite sequence $x_k$ with $k = 1, \dots, n$ and $x_k \in [a,b]$, for any $\epsilon > 0$ the value $y_k = x_k - a + \epsilon$ belongs to $[\epsilon, \rho + \epsilon]$, where $\rho = b - a$. Since

$$\overline{y} = \overline{x} - a + \epsilon \,,$$

we have $y_k - \overline{y} = x_k - \overline{x}$ for every value of $k$. Therefore, by applying the previous theorem to the sequence $y_k$, we have that

$$\frac{1}{2(\rho+\epsilon)}\sum_{k=1}^{n} p_k(x_k-\overline{x})^2 = \frac{1}{2(\rho+\epsilon)}\sum_{k=1}^{n} p_k(y_k-\overline{y})^2 \leq \overline{y} - \prod_{k=1}^{n} y_k^{p_k} .$$

By considering the limit $\epsilon \to 0$, since $\rho \geq r > 0$ then it follows that

$$\frac{1}{2\rho}\sum_{k=1}^{n} p_k(x_k-\overline{x})^2 \leq \overline{x} - a - \prod_{k=1}^{n}(x_k-a)_k^{p_k} .$$

However, if we set $a = \min_k x_k$ then $x_k - a = 0$ for some $k = 1, \dots, n$ and hence

$$\frac{1}{2\rho}\sum_{k=1}^{n} p_k(x_k-\overline{x})^2 \leq \overline{x} - \min_k x_k .$$

By repeating the steps above with the sequence $y_k = \epsilon + b - x_k$, we get

$$\frac{1}{2\rho}\sum_{k=1}^{n} p_k(x_k-\overline{x})^2 \leq \max_k x_k - \overline{x} .$$

Since for the selected values of $a$ and $b$ we have that $\rho = r$, the thesis follows by combining the last two inequalities. ■

*Corollary:* If $X$ is a random variable with finite support $\{x_1, \dots, x_n\}$ and probability distribution given by $p_k = P(X = x_k)$, then

$$\min(\mathrm{X}) \leq E[X] \pm \frac{Var(X)}{2Ran(X)} \leq \max(X) .$$

Furthermore, even though we will not deepen this topic in this work, by using classic approximation theorems this result can be extended to any bounded random variable on any probability space.

*Remark:* Given $\epsilon > 0$, consider a finite sequence $x_k^\epsilon$, with $k = 1, \dots, n$. Assume $x_k^\epsilon$ to be not constant for every $\epsilon > 0$, and denote by $\overline{x}_\epsilon$ and $r_\epsilon$ its weighted mean and range, respectively. If $x_k^\epsilon \to \overline{x}$ as $\epsilon \to 0$ for every $k = 1, \dots, n$, then by the formula $(A2)$ and the squeeze theorem we have

$$\overline{x}_\epsilon \pm \frac{1}{2r_\epsilon}\sum_{k=1}^{n} p_k(x_k^\epsilon - \overline{x}_\epsilon)^2 \to \overline{x} , \qquad \epsilon \to 0 .$$

*Definition:* Given a finite sequence $x_k$, with $k = 1, \dots, n$, we define its *penalized arithmetic mean for positive/negative polarity* as

$$\overline{x}_{\pm} = \overline{x} \mp \frac{var(x)}{2ran(x)},$$

with $\overline{x}_{\pm} = \overline{x}$ whether $x_k$ is constant. More in general, given a bounded random variable $X$, we will define its *penalized average for positive/negative polarity* as

$$E[X]_{\pm} = E[X] \mp \frac{Var(X)}{2Ran(X)},$$

with $E[X]_{\pm} = E[X]$ by definition whether $X$ is constant.

*Remark:* Given $\lambda > 0$ and $c \in \mathbb{R}$, by the basic properties of mean, variance, and range, it is trivial to notice that

$$\overline{\lambda x}_{\pm} = \lambda \overline{x}_{\pm}\,, \qquad \overline{x + c}_{\pm} = \overline{x}_{\pm} + c\,, \qquad \overline{-x}_{\pm} = -\overline{x}_{\mp}\,.$$

In addition, by using these properties it can be easily shown that if $x$ consists of two values $x_0 < x_1$ then its penalized averages are equal to

$$\overline{x}_{+} = \frac{5x_0 + 3x_1}{8}\,, \qquad \overline{x}_{-} = \frac{3x_0 + 5x_1}{8}\,.$$

These penalized means are an attempt to define an aggregation method that is sensitive to small values, like the geometric mean, but mathematically more similar to the arithmetic mean, so that it does not vanish when one of the values is zero. This approach is particularly useful for composite indicators, especially when compensation between indicators from different domains is undesirable and when imbalance among the indicators should be penalized.

**Table A1** – *Comparison between the arithmetic mean $\overline{\boldsymbol{x}}$, the penalized arithmetic mean $\overline{\boldsymbol{x}}_{+}$, and the geometric mean $\overline{\boldsymbol{x}}_{\boldsymbol{g}}$ for different sequences.*

| Sequence | $\overline{x}$ | $\overline{x}_{+}$ | $\overline{x}_{g}$ |
|---|---|---|---|
| 0,0,0,0,0,0,0,0,0,10 | 1 | 0.55 | 0 |
| 1,1,1,1,1,1,1,1,1,91 | 10 | 5.95 | 1.57 |
| 4,6 | 5 | 4.75 | 4.90 |
| 2,8 | 5 | 4.25 | 4 |
| $-1$,1 | 0 | $-0.25$ | Not defined |

We can notice from Table A1 that the penalized arithmetic mean can be smaller or larger than the geometric mean. More specifically, it results to be smaller than the geometric mean when the values being compared are relatively close to each other.

This behaviour can be useful when working with standardized data, where extreme outliers have been minimized. However, when using these means to aggregate composite indicators, due to this fact it may be argued that the penalized arithmetic mean does not align well with the geometric mean. In other words, due to the lack of some penalized arithmetic mean-geometric mean inequality, experts in the field should carefully decide whether to use these two means for ensuring a gradually less compensatory approach between different domains.

## 9. Annexes

**Annex 1 –** *Descriptive of the final indicators used in calculating the IGEI*

| Indicator | mean | sd | cv | min | p25 | p50 | p75 | max |
|---|---|---|---|---|---|---|---|---|
| 1 | 72.26 | 13.32 | 0.18 | 44.05 | 66.74 | 78.77 | 79.92 | 88.65 |
| 2 | 87.52 | 3.59 | 0.04 | 80.23 | 85.13 | 87.05 | 90.30 | 95.94 |
| 3 | 38.27 | 8.19 | 0.21 | 25.87 | 32.99 | 36.28 | 43.17 | 53.45 |
| 4 | 58.21 | 6.04 | 0.10 | 47.63 | 53.38 | 58.48 | 61.75 | 70.46 |
| 5 | 71.13 | 2.20 | 0.03 | 66.97 | 69.83 | 71.43 | 72.61 | 74.43 |
| 6 | 57.67 | 7.45 | 0.13 | 40.88 | 52.21 | 58.34 | 64.20 | 67.34 |
| 7 | 77.09 | 10.83 | 0.14 | 57.74 | 66.47 | 80.98 | 85.43 | 92.05 |
| 8 | 28.49 | 6.27 | 0.22 | 9.59 | 24.74 | 29.98 | 33.60 | 37.00 |
| 9 | 59.61 | 13.05 | 0.22 | 32.74 | 52.30 | 58.79 | 69.32 | 86.16 |
| 10 | 71.68 | 11.49 | 0.16 | 43.67 | 66.53 | 74.53 | 79.40 | 90.07 |
| 11 | 32.65 | 9.21 | 0.28 | 13.90 | 25.95 | 33.80 | 39.50 | 46.50 |
| 12 | 75.49 | 6.05 | 0.08 | 65.11 | 71.87 | 75.65 | 79.23 | 85.15 |
| 13 | 29.09 | 5.60 | 0.19 | 17.47 | 25.18 | 29.38 | 32.86 | 41.75 |
| 14 | 58.98 | 22.41 | 0.38 | 33.33 | 38.12 | 56.49 | 70.25 | 100.00 |
| 15 | 40.84 | 21.86 | 0.54 | 12.87 | 23.84 | 39.08 | 55.40 | 90.84 |
| 16 | 52.42 | 9.00 | 0.17 | 37.17 | 47.17 | 51.98 | 59.63 | 67.79 |
| 17 | 94.50 | 0.77 | 0.01 | 92.85 | 94.20 | 94.58 | 95.06 | 95.44 |
| 18 | 86.53 | 3.55 | 0.04 | 81.32 | 83.78 | 86.21 | 87.81 | 95.93 |
| 19 | 90.51 | 1.68 | 0.02 | 88.65 | 89.34 | 90.14 | 91.07 | 95.41 |
| 20 | 90.47 | 3.44 | 0.04 | 84.09 | 87.10 | 91.12 | 92.72 | 96.39 |
| 21 | 80.72 | 5.24 | 0.06 | 73.11 | 76.77 | 80.53 | 84.39 | 91.35 |
| 22 | 62.68 | 11.63 | 0.19 | 36.84 | 53.37 | 66.50 | 72.01 | 77.91 |

**Annex 2 –** *Correlation matrix of the 22 gender indicators*

| | 1 | 2 | 3 | 4 | 5 | 6 | 7 | 8 | 9 | 10 | 11 | 12 | 13 | 14 | 15 | 16 | 17 | 18 | 19 | 20 | 21 | 22 |
|---|---|---|---|---|---|---|---|---|---|---|---|---|---|---|---|---|---|---|---|---|---|---|
| 1 | 1.00 | | | | | | | | | | | | | | | | | | | | | |
| 2 | 0.61 | 1.00 | | | | | | | | | | | | | | | | | | | | |
| 3 | -0.89 | -0.72 | 1.00 | | | | | | | | | | | | | | | | | | | |
| 4 | 0.71 | 0.47 | -0.74 | 1.00 | | | | | | | | | | | | | | | | | | |
| 5 | 0.26 | -0.05 | -0.10 | 0.24 | 1.00 | | | | | | | | | | | | | | | | | |
| 6 | 0.13 | -0.05 | -0.23 | 0.42 | 0.16 | 1.00 | | | | | | | | | | | | | | | | |
| 7 | 0.83 | 0.49 | -0.85 | 0.71 | 0.09 | 0.24 | 1.00 | | | | | | | | | | | | | | | |
| 8 | 0.29 | -0.03 | -0.27 | 0.36 | 0.08 | 0.43 | 0.41 | 1.00 | | | | | | | | | | | | | | |
| 9 | -0.09 | -0.42 | 0.27 | 0.00 | 0.49 | 0.46 | -0.15 | 0.21 | 1.00 | | | | | | | | | | | | | |
| 10 | 0.85 | 0.38 | -0.64 | 0.54 | 0.32 | -0.01 | 0.67 | 0.41 | 0.15 | 1.00 | | | | | | | | | | | | |
| 11 | 0.85 | 0.26 | -0.73 | 0.65 | 0.39 | 0.29 | 0.83 | 0.55 | 0.24 | 0.86 | 1.00 | | | | | | | | | | | |
| 12 | 0.70 | 0.56 | -0.58 | 0.45 | 0.14 | -0.41 | 0.50 | 0.12 | -0.35 | 0.70 | 0.48 | 1.00 | | | | | | | | | | |
| 13 | 0.85 | 0.47 | -0.77 | 0.50 | 0.19 | 0.07 | 0.83 | 0.30 | -0.13 | 0.80 | 0.79 | 0.58 | 1.00 | | | | | | | | | |
| 14 | 0.39 | 0.56 | -0.45 | 0.11 | 0.17 | 0.11 | 0.41 | 0.02 | -0.23 | 0.11 | 0.27 | 0.17 | 0.33 | 1.00 | | | | | | | | |
| 15 | 0.50 | 0.23 | -0.52 | 0.62 | 0.17 | 0.57 | 0.59 | 0.50 | 0.20 | 0.52 | 0.59 | 0.26 | 0.50 | -0.01 | 1.00 | | | | | | | |
| 16 | 0.44 | 0.06 | -0.40 | 0.41 | 0.22 | 0.33 | 0.44 | 0.31 | 0.19 | 0.39 | 0.58 | 0.13 | 0.34 | 0.45 | 0.17 | 1.00 | | | | | | |
| 17 | 0.75 | 0.63 | -0.74 | 0.68 | 0.05 | 0.19 | 0.65 | 0.13 | -0.03 | 0.62 | 0.62 | 0.64 | 0.59 | 0.31 | 0.59 | 0.41 | 1.00 | | | | | |
| 18 | 0.47 | 0.56 | -0.49 | 0.45 | -0.24 | -0.16 | 0.49 | 0.00 | -0.59 | 0.31 | 0.15 | 0.47 | 0.42 | 0.12 | 0.29 | -0.16 | 0.35 | 1.00 | | | | |
| 19 | 0.07 | 0.21 | -0.17 | -0.17 | -0.28 | -0.12 | 0.04 | -0.35 | -0.57 | -0.17 | -0.20 | 0.08 | 0.12 | 0.31 | 0.02 | -0.13 | 0.07 | 0.49 | 1.00 | | | |
| 20 | 0.59 | 0.62 | -0.64 | 0.45 | -0.15 | 0.10 | 0.66 | 0.35 | -0.34 | 0.40 | 0.38 | 0.50 | 0.47 | 0.39 | 0.50 | 0.07 | 0.42 | 0.70 | 0.32 | 1.00 | | |
| 21 | -0.67 | -0.33 | 0.54 | -0.27 | -0.23 | 0.14 | -0.62 | -0.13 | 0.26 | -0.60 | -0.58 | -0.45 | -0.82 | -0.35 | -0.12 | -0.11 | -0.21 | -0.44 | -0.19 | -0.31 | 1.00 | |
| 22 | 0.09 | -0.15 | -0.09 | 0.06 | 0.50 | 0.26 | 0.13 | 0.36 | 0.15 | 0.01 | 0.25 | -0.14 | 0.14 | 0.15 | 0.07 | 0.27 | -0.11 | -0.34 | -0.33 | -0.09 | -0.04 | 1.00 |

**Annex 3.1 –** *Indicator values for the domains work, economy and knowledge*

| Region | Work | | | Money | | Knowledge | | | |
|---|---|---|---|---|---|---|---|---|---|
| | 1 | 2 | 3 | 4 | 5 | 6 | 7 | 8 | 9 |
| Emilia-Romagna | 79.38 | 90.21 | 32.95 | 63.60 | 74.43 | 65.52 | 86.04 | 33.64 | 64.30 |
| Umbria | 78.77 | 84.60 | 38.40 | 58.52 | 72.29 | 64.21 | 85.56 | 33.57 | 86.16 |
| Friuli-Venezia Giulia | 81.89 | 89.07 | 34.12 | 60.41 | 70.88 | 55.80 | 91.61 | 28.37 | 51.95 |
| Toscana | 79.23 | 86.50 | 36.37 | 62.09 | 71.85 | 61.63 | 78.77 | 32.51 | 70.70 |
| Veneto | 78.09 | 90.65 | 32.05 | 61.42 | 67.01 | 64.82 | 91.45 | 29.98 | 62.06 |
| Provincia Autonoma di Trento | 83.77 | 90.39 | 25.87 | 56.04 | 68.32 | 53.57 | 92.05 | 37.00 | 33.73 |
| Lazio | 71.23 | 84.87 | 38.34 | 68.31 | 72.84 | 64.68 | 85.29 | 34.91 | 66.93 |
| Piemonte | 79.34 | 88.99 | 36.12 | 64.69 | 73.27 | 64.18 | 76.59 | 30.29 | 55.34 |
| Trentino-Alto Adige/Südtirol | 84.97 | 93.25 | 26.48 | 58.49 | 68.96 | 52.12 | 87.42 | 24.20 | 31.78 |
| Valle d'Aosta/Vallée d'Aoste | 88.65 | 90.88 | 33.04 | 55.83 | 73.93 | 44.46 | 81.23 | 23.00 | N.A. |
| Lombardia | 78.84 | 90.56 | 29.48 | 70.46 | 71.43 | 58.34 | 81.30 | 30.28 | 56.82 |
| Marche | 80.47 | 89.05 | 36.28 | 58.48 | 71.72 | 58.08 | 66.72 | 22.70 | 73.89 |
| **Italia** | **68.23** | **87.92** | **37.96** | **63.00** | **71.00** | **59.74** | **77.35** | **27.99** | **58.49** |
| Provincia Autonoma di Bolzano/Bozen | 86.25 | 95.94 | 27.12 | 60.78 | 69.64 | 50.55 | 81.74 | 9.59 | 32.74 |
| Abruzzo | 65.92 | 86.15 | 46.65 | 52.35 | 70.01 | 67.34 | 66.22 | 33.90 | 73.71 |
| Liguria | 76.00 | 85.39 | 34.76 | 58.27 | 70.53 | 50.86 | 80.98 | 28.72 | 57.13 |
| Sardegna | 67.56 | 80.23 | 39.68 | 54.42 | 72.38 | 64.08 | 75.31 | 26.53 | 60.45 |
| Calabria | 44.05 | 87.05 | 49.28 | 47.63 | 74.41 | 57.67 | 59.37 | 22.31 | 70.12 |
| Molise | 60.66 | 86.06 | 52.99 | 51.08 | 70.44 | 40.88 | 64.38 | 34.45 | 53.35 |
| Sicilia | 45.74 | 81.64 | 53.45 | 50.98 | 69.09 | 60.14 | 57.74 | 25.45 | 51.67 |
| Puglia | 47.13 | 84.70 | 50.18 | 50.61 | 66.97 | 48.97 | 62.36 | 24.03 | 51.90 |
| Basilicata | 53.54 | 79.77 | 58.69 | 49.59 | 72.20 | 46.11 | 64.66 | 11.89 | 68.16 |
| Campania | 41.72 | 86.95 | 51.49 | 51.22 | 67.49 | 55.92 | 61.05 | 18.36 | 54.96 |

NB: The N.A. cell in the last column is due to the lack of universities in the Valle d'Aosta region with faculties considered by the corresponding indicator

**Annex 3.2 –** *Indicator values for the domains time, politics and health*

| Region | Time | | | | Politics | | | Health | | | | | |
|---|---|---|---|---|---|---|---|---|---|---|---|---|---|
| | 10 | 11 | 12 | 13 | 14 | 15 | 16 | 17 | 18 | 19 | 20 | 21 | 22 |
| Emilia-Romagna | 77.29 | 43.10 | 78.25 | 30.13 | 95.31 | 56.25 | 64.74 | 95.41 | 85.68 | 90.40 | 93.22 | 83.06 | 68.80 |
| Umbria | 90.07 | 46.50 | 72.93 | 33.97 | 49.93 | 90.84 | 51.98 | 95.01 | 86.23 | 91.01 | 91.16 | 78.99 | 46.50 |
| Friuli-Venezia Giulia | 74.53 | 38.30 | 70.81 | 34.93 | 100.00 | 23.61 | 67.79 | 94.38 | 91.68 | 91.13 | 91.74 | 73.11 | 72.08 |
| Toscana | 78.72 | 40.70 | 79.34 | 28.91 | 56.49 | 53.85 | 63.40 | 95.30 | 85.90 | 89.30 | 91.12 | 82.58 | 66.50 |
| Veneto | 73.25 | 33.80 | 75.53 | 34.14 | 67.79 | 54.56 | 58.48 | 94.97 | 86.99 | 89.56 | 96.36 | 81.58 | 52.71 |
| Provincia Autonoma di Trento | 80.09 | 41.20 | 81.37 | 41.75 | 62.50 | 66.67 | 47.06 | 95.05 | 89.45 | 92.75 | 94.66 | 73.26 | 75.75 |
| Lazio | 67.14 | 37.90 | 74.30 | 26.60 | 37.55 | 70.07 | 47.28 | 94.52 | 87.51 | 88.65 | 91.86 | 82.78 | 73.08 |
| Piemonte | 81.51 | 32.70 | 77.01 | 31.33 | 38.70 | 70.07 | 50.15 | 94.58 | 87.73 | 88.83 | 90.28 | 80.37 | 74.23 |
| Trentino-Alto Adige/Südtirol | 73.09 | 31.60 | 83.10 | 37.13 | 85.87 | 52.21 | 42.65 | 95.39 | 92.69 | 94.09 | 95.53 | 74.14 | 58.23 |
| Valle d'Aosta/Vallée d'Aoste | 86.21 | 43.00 | 84.94 | 33.59 | 100.00 | 12.87 | 61.81 | 94.83 | 81.32 | 88.77 | 88.23 | 73.73 | 68.16 |
| Lombardia | 75.42 | 36.00 | 78.96 | 32.12 | 40.65 | 39.08 | 54.56 | 95.13 | 87.88 | 89.21 | 86.92 | 78.54 | 54.03 |
| Marche | 81.78 | 33.50 | 75.65 | 29.19 | 36.43 | 40.85 | 57.73 | 95.08 | 86.02 | 90.87 | 88.81 | 85.71 | 62.93 |
| **Italia** | **65.87** | **30.00** | **73.92** | **28.89** | **50.83** | **35.87** | **51.75** | **94.47** | **86.06** | **90.00** | **89.55** | **83.06** | **61.83** |
| Provincia Autonoma di Bolzano/Bozen | 67.60 | 23.90 | 85.15 | 31.38 | 85.71 | 40.06 | 37.17 | 95.44 | 95.93 | 95.41 | 96.39 | 75.00 | 36.84 |
| Abruzzo | 61.32 | 28.00 | 66.09 | 25.26 | 62.60 | 24.07 | 42.05 | 94.13 | 83.15 | 89.38 | 87.04 | 80.53 | 67.75 |
| Liguria | 70.70 | 33.80 | 79.11 | 24.83 | 36.43 | 29.20 | 49.25 | 94.53 | 86.21 | 89.80 | 92.22 | 86.47 | 71.95 |
| Sardegna | 65.92 | 35.20 | 65.11 | 29.38 | 45.56 | 20.05 | 55.04 | 92.85 | 83.67 | 91.50 | 86.92 | 73.86 | 65.66 |
| Calabria | 48.91 | 15.70 | 66.22 | 22.26 | 72.71 | 24.07 | 40.45 | 93.21 | 81.41 | 90.14 | 86.78 | 86.41 | 77.91 |
| Molise | 75.54 | 22.50 | 82.11 | 25.10 | 33.33 | 16.69 | 37.17 | 93.00 | 90.55 | 90.28 | 94.01 | 78.55 | 50.50 |
| Sicilia | 43.67 | 13.90 | 67.03 | 17.47 | 65.56 | 27.23 | 60.77 | 93.76 | 82.81 | 93.13 | 87.16 | 91.35 | 62.53 |
| Puglia | 62.32 | 20.60 | 74.33 | 20.45 | 33.33 | 15.87 | 49.03 | 94.28 | 83.90 | 89.60 | 84.09 | 87.79 | 43.08 |
| Basilicata | 63.57 | 22.40 | 72.47 | 27.97 | 16.69 | 16.69 | 42.25 | 93.99 | 83.86 | 89.86 | 81.76 | 80.38 | 62.81 |
| Campania | 44.83 | 13.20 | 58.94 | 15.43 | 42.05 | 18.62 | 38.12 | 93.39 | 85.43 | 90.04 | 88.47 | 90.77 | 54.26 |

**Annex 4.1 –** *IGEI domains and sub-domains scores (work, money and knowledge)*

| Region | IGEI | Work | Participation | Entrepreneurship and quality of work | Money | Labour earnings | Pension income | Knowledge | Attainment and participation | Segregation |
|---|---|---|---|---|---|---|---|---|---|---|
| Emilia-Romagna | 66.76 | 68.25 | 79.38 | 61.58 | 67.66 | 63.60 | 74.43 | 59.02 | 75.78 | 48.97 |
| Umbria | 64.39 | 67.98 | 78.77 | 61.50 | 63.69 | 58.52 | 72.29 | 65.50 | 74.89 | 59.86 |
| Friuli-Venezia Giulia | 63.37 | 69.20 | 81.89 | 61.60 | 64.34 | 60.41 | 70.88 | 52.74 | 73.70 | 40.16 |
| Toscana | 63.33 | 68.11 | 79.23 | 61.43 | 65.75 | 62.09 | 71.85 | 58.58 | 70.20 | 51.60 |
| Veneto | 62.89 | 67.63 | 78.09 | 61.35 | 63.52 | 61.42 | 67.01 | 58.06 | 78.13 | 46.02 |
| Provincia Autonoma di Trento | 62.17 | 67.74 | 83.77 | 58.13 | 60.64 | 56.04 | 68.32 | 49.41 | 72.81 | 35.37 |
| Lazio | 60.77 | 65.21 | 71.23 | 61.61 | 70.01 | 68.31 | 72.84 | 59.95 | 74.99 | 50.92 |
| Piemonte | 60.71 | 68.85 | 79.34 | 62.56 | 67.91 | 64.69 | 73.27 | 53.15 | 70.38 | 42.81 |
| Trentino-Alto Adige/Südtirol | 60.69 | 69.28 | 84.97 | 59.87 | 62.41 | 58.49 | 68.96 | 43.66 | 69.77 | 27.99 |
| Valle d'Aosta/Vallée d'Aoste | 60.60 | 71.97 | 88.65 | 61.96 | 62.62 | 55.83 | 73.93 | 37.94 | 62.85 | 23.00 |
| Lombardia | 59.58 | 67.08 | 78.84 | 60.02 | 70.82 | 70.46 | 71.43 | 53.40 | 69.82 | 43.55 |
| Marche | 58.80 | 69.34 | 80.47 | 62.67 | 63.44 | 58.48 | 71.72 | 53.58 | 62.40 | 48.29 |
| **Italia** | **58.20** | **64.92** | **68.23** | **62.94** | **66.00** | **63.00** | **71.00** | **52.73** | **68.54** | **43.24** |
| Provincia Autonoma di Bolzano/Bozen | 57.61 | 70.80 | 86.25 | 61.53 | 64.10 | 60.78 | 69.64 | 38.03 | 66.14 | 21.16 |
| Abruzzo | 57.01 | 66.10 | 65.92 | 66.40 | 58.97 | 52.35 | 70.01 | 58.67 | 66.78 | 53.81 |
| Liguria | 57.01 | 66.05 | 76.00 | 60.08 | 62.87 | 58.27 | 70.53 | 51.55 | 65.92 | 42.93 |
| Sardegna | 55.83 | 62.81 | 67.56 | 59.95 | 61.15 | 54.42 | 72.38 | 53.32 | 69.69 | 43.49 |
| Calabria | 52.90 | 53.09 | 44.05 | 68.16 | 57.67 | 47.63 | 74.41 | 50.83 | 58.52 | 46.22 |
| Molise | 52.63 | 63.99 | 60.66 | 69.53 | 58.34 | 51.08 | 70.44 | 47.18 | 52.63 | 43.90 |
| Sicilia | 52.33 | 53.91 | 45.74 | 67.55 | 57.77 | 50.98 | 69.09 | 46.20 | 58.94 | 38.56 |
| Puglia | 49.58 | 54.74 | 47.13 | 67.44 | 56.75 | 50.61 | 66.97 | 44.60 | 55.67 | 37.96 |
| Basilicata | 49.10 | 59.42 | 53.54 | 69.23 | 58.07 | 49.59 | 72.20 | 45.79 | 55.38 | 40.03 |
| Campania | 47.49 | 52.03 | 41.72 | 69.22 | 57.32 | 51.22 | 67.49 | 44.84 | 58.49 | 36.66 |

**Annex 4.2 –** *IGEI domains and sub-domains scores (time, politics and health)*

| Region | IGEI | Time | Care activities | Social activities | Politics | National representation | Subnational representation | Health | Status | Behaviour |
|---|---|---|---|---|---|---|---|---|---|---|
| Emilia-Romagna | 66.76 | 56.44 | 60.20 | 54.19 | 73.55 | 95.31 | 60.50 | 85.00 | 90.50 | 81.69 |
| Umbria | 64.39 | 59.01 | 68.28 | 53.45 | 57.98 | 49.93 | 71.41 | 79.17 | 90.75 | 72.22 |
| Friuli-Venezia Giulia | 63.37 | 54.20 | 56.41 | 52.87 | 66.06 | 100.00 | 45.70 | 84.01 | 92.39 | 78.98 |
| Toscana | 63.33 | 56.22 | 59.71 | 54.12 | 57.29 | 56.49 | 58.62 | 83.85 | 90.17 | 80.06 |
| Veneto | 62.89 | 54.02 | 53.53 | 54.83 | 60.74 | 67.79 | 56.52 | 81.99 | 90.50 | 76.88 |
| Provincia Autonoma di Trento | 62.17 | 60.99 | 60.65 | 61.56 | 58.98 | 62.50 | 56.86 | 85.42 | 92.42 | 81.22 |
| Lazio | 60.77 | 51.22 | 52.52 | 50.45 | 45.47 | 37.55 | 58.67 | 85.44 | 90.23 | 82.57 |
| Piemonte | 60.71 | 55.27 | 57.10 | 54.17 | 46.73 | 38.70 | 60.11 | 84.91 | 90.38 | 81.63 |
| Trentino-Alto Adige/Südtirol | 60.69 | 55.26 | 52.35 | 60.11 | 61.85 | 85.87 | 47.43 | 82.75 | 94.06 | 75.97 |
| Valle d'Aosta/Vallée d'Aoste | 60.60 | 61.27 | 64.60 | 59.27 | 60.84 | 100.00 | 37.34 | 81.06 | 88.31 | 76.71 |
| Lombardia | 59.58 | 55.60 | 55.71 | 55.54 | 42.96 | 40.65 | 46.82 | 79.75 | 90.74 | 73.16 |
| Marche | 58.80 | 54.38 | 57.64 | 52.42 | 41.25 | 36.43 | 49.29 | 83.46 | 90.65 | 79.15 |
| Italia | **58.20** | **49.24** | **47.94** | **51.41** | **46.44** | **50.83** | **43.81** | **82.66** | **90.17** | **78.15** |
| Provincia Autonoma di Bolzano/Bozen | 57.61 | 50.44 | 45.75 | 58.26 | 56.28 | 85.71 | 38.62 | 79.23 | 95.59 | 69.41 |
| Abruzzo | 57.01 | 45.04 | 44.66 | 45.68 | 44.14 | 62.60 | 33.06 | 82.36 | 88.89 | 78.44 |
| Liguria | 57.01 | 52.08 | 52.25 | 51.97 | 37.48 | 36.43 | 39.23 | 86.03 | 90.18 | 83.55 |
| Sardegna | 55.83 | 48.49 | 50.56 | 47.25 | 40.55 | 45.56 | 37.54 | 80.68 | 89.34 | 75.48 |
| Calabria | 52.90 | 36.78 | 32.30 | 44.24 | 47.43 | 72.71 | 32.26 | 85.41 | 88.25 | 83.70 |
| Molise | 52.63 | 50.74 | 49.02 | 53.61 | 29.33 | 33.33 | 26.93 | 80.70 | 91.28 | 74.35 |
| Sicilia | 52.33 | 33.83 | 28.79 | 42.25 | 52.09 | 65.56 | 44.00 | 83.93 | 89.90 | 80.35 |
| Puglia | 49.58 | 43.69 | 41.46 | 47.39 | 32.78 | 33.33 | 32.45 | 78.26 | 89.26 | 71.65 |
| Basilicata | 49.10 | 45.70 | 42.99 | 50.22 | 21.48 | 16.69 | 29.47 | 80.33 | 89.24 | 74.98 |
| Campania | 47.49 | 32.08 | 29.01 | 37.18 | 33.50 | 42.05 | 28.37 | 82.25 | 89.62 | 77.83 |

**Annex 5.1 –** *Indicator details for the domains work, economy, knowledge, and time*

| Indicator | Definition | Source |
|---|---|---|
| Employment rate | Percentage of employed persons aged 20–64 over the total population aged 20–64. | BES (Istat - Rilevazione sulle Forze di lavoro) |
| Entrepreneurial density | Gender-specific entrepreneurial density, defined as the ratio of female-owned enterprises to the female labour force and male-owned enterprises to the male labour force, both aged 15–64. Registered enterprises as of 31 December 2024; labour force measured in 2024-Q4 (last three months of 2024). | Movimprese e Osservatorio Imprenditorialità Femminile; Unioncamere - Infocamere; Istat - Rilevazione sulle Forze di lavoro |
| Involuntary part-time employment | Percentage of employed persons who report working part-time because they could not find a full-time job, out of total employment. | BES (Istat - Rilevazione sulle Forze di lavoro) |
| Average annual gross wage of employees | Ratio between the total annual gross earnings (before personal income tax) of private-sector non-agricultural employees insured with INPS and the number of employees (in euros). | BES dei Territori (Elaborazioni su dati Inps - Osservatorio sui lavoratori dipendenti) |
| Average annual per-capita pension income | Ratio between the total amount of pensions paid during the year (in euros) and the number of pensioners. | BES dei Territori (Statistiche della previdenza e dell'assistenza sociale) |
| Tertiary graduates and other tertiary qualifications | Percentage of individuals aged 25–34 who have obtained a tertiary-level qualification (ISCED 5, 6, 7, or 8) over the total population aged 25–34. | BES (Istat - Rilevazione sulle Forze di lavoro) |
| Participation in lifelong learning | Percentage of individuals aged 25–64 who participated in education or training activities in the four weeks preceding the interview, over the total population aged 25–64. | BES (Istat - Rilevazione sulle Forze di lavoro) |
| Graduates in female-dominated fields | Graduates in Arts and Humanities, Education, Health and Welfare (ISCED-F classification). | Ministero dell'Università e della Ricerca |
| Graduates in male-dominated fields | Graduates in Engineering, Manufacturing and Construction, Information and Communication Technologies (ICT), Natural Sciences, Mathematics and Statistics (ISCED-F classification). | Ministero dell'Università e della Ricerca |
| Ratio between employment rates of women with preschool children and women without children | Employment rate of women aged 25–49 with at least one child aged 0–5 divided by the employment rate of women aged 25–49 without children, multiplied by 100. | BES (Istat - Rilevazione sulle Forze di lavoro) |
| Authorized places in early childhood education and care services | Available places as of 31 December 2022 in early childhood education services per 100 children aged 0–2. This includes traditional childcare centres, micro-nurseries, workplace nurseries, spring sections, and supplementary early childhood services. | Istat - I servizi educativi per l'infanzia in Italia |
| Social participation | Percentage of individuals aged 14 and over who, in the previous 12 months, engaged in at least one social participation activity, over the total population aged 14 and over. Activities include participation in cultural, sports, recreational, or spiritual initiatives promoted by religious or spiritual groups; meetings of cultural, recreational, environmental, civil rights, peace, trade union, professional, or political associations; unpaid activities for political parties; or payment of membership fees for sports clubs. | BES (Istat - Indagine Aspetti della vita quotidiana) |
| Volunteering activities | Percentage of individuals aged 14 and over who, in the previous 12 months, performed unpaid work for voluntary associations or groups, over the total population aged 14 and over. | BES (Istat - Indagine Aspetti della vita quotidiana) |

**Annex 5.2** – *Indicator details for the domains politics and health*

| Indicator | Definition | Source |
| --- | --- | --- |
| **Women and political representation in Parliament** | Percentage of women elected to the Senate of the Republic and the Chamber of Deputies over the total number of elected representatives. Senators and deputies elected in overseas constituencies and life senators are excluded. | BES (Istat - Elaborazione su dati della Camera dei Deputati e del Senato della Repubblica) |
| **Women and political representation at the local level** | Percentage of women elected to Regional Councils over the total number of elected representatives. | BES (Istat - Elaborazione su dati dei Consigli regionali) |
| **Women in municipal administrators** | Percentage of women over the total number of municipally elected administrators. | BES dei Territori (Elaborazione su dati Ministero dell'Interno - Anagrafe degli amministratori locali) |
| **Life expectancy at birth** | Life expectancy expresses the average number of years a newborn in a given calendar year can expect to live. | BES (Istat - Tavole di mortalità della popolazione italiana) |
| **Healthy life expectancy at birth** | Average number of years a newborn can expect to live in good health, based on the prevalence of individuals reporting “good” or “very good” perceived health. | BES (Istat - Tavole di mortalità della popolazione italiana e Indagine Aspetti della vita quotidiana) |
| **Mental health** | Measure of psychological distress derived from the synthesis of scores obtained by individuals aged 14 and over on five questions from the SF-36 questionnaire. The index refers to four main dimensions of mental health (anxiety, depression, loss of behavioural or emotional control, and psychological well-being). The standardized index ranges from 0 to 100, with higher values indicating better mental well-being. | BES (Istat - Indagine Aspetti della vita quotidiana) |
| **Smoking** | Proportion of individuals aged 14 and over who report currently smoking, standardized to the 2013 European population, over the total population aged 14 and over. | BES (Istat - Indagine Aspetti della vita quotidiana) |
| **Alcohol** | Proportion of individuals aged 14 and over exhibiting at least one risky alcohol consumption behaviour, standardized to the 2013 European population. Risky consumption is defined according to Ministry of Health guidelines (LARN 2014) and criteria established with the National Institute of Health, including exceeding daily consumption thresholds (by sex and age) or consuming six or more alcohol units on a single occasion (binge drinking). | BES (Istat - Indagine Aspetti della vita quotidiana) |
| **Adequate nutrition** | Proportion of individuals aged 3 and over who consume at least four portions of fruit and/or vegetables daily, standardized to the 2013 European population, over the total population aged 3 and over. | BES (Istat - Indagine Aspetti della vita quotidiana) |

## 10.References